\documentclass[pra,twocolumn,showpacs]{revtex4}
\usepackage{graphicx}

\begin{document}
\title{Generation of entangled coherent states for distant Bose-Einstein condensates via electromagnetically induced transparency}
\author{Le-Man Kuang$^{1,2}$, Zeng-Bing Chen$^{2,3}$, and Jian-Wei Pan$^{2,3}$}
\affiliation{$^{1}$Key Laboratory of Low-Dimensional Quantum Structures and Quantum Control of Ministry of Education, and Department of Physics, Hunan Normal University, Changsha 410081, China\\
$^{2}$ Physikalisches Institut, Universit\"{a}t Heidelberg,
Philosophenweg 12, 69120 Heidelberg, Germany\\
$^{3}$Hefei National Laboratory for Physical Sciences at
Microscale and Department of Modern Physics, University of Science
and Technology of China, Hefei 230026, China}
\begin{abstract}
In this paper, we propose a method to generate entangled coherent
states between two spatially separated atomic Bose-Einstein
condensates (BECs) via the technique of the electromagnetically
induced transparency (EIT).  Two strong coupling laser beams and
two entangled probe laser beams are used to make two distant BECs
be in EIT states and to generate an atom-photon entangled state
between probe lasers and distant BECs. The two BECs are initially
in un-entangled product coherent states while the probe lasers are
initially in an entangled state. Entangled states of two distant
BECs can be created through performing projective measurements
upon the two outgoing probe lasers under certain conditions.
Concretely, we propose two protocols to show how to generate
entangled coherent states of the two distant BECs. One is a
single-photon scheme in which an entangled single-photon state
serves the quantum channel to generate entangled distant BECs. The
other is a multiphoton scheme where an entangled coherent state of
the probe lasers is used as the quantum channel. Additionally, we
also obtain some atom-photon entangled states of particular
interest such as entangled states between a pair of optical Bell
(or quasi-Bell) states and a pair of atomic entangled coherent (or
quasi-Bell) states.
\end{abstract}

 \maketitle

\section{Introduction}
According to quantum state superposition principle in quantum mechanics \cite{per}, a quantum
system may exist at once in several eigenstates corresponding to different eigenvalues of a
physical observable. Quantum entanglement is a direct consequence of this principle applied to a
composite quantum system. An essential feature of quantum entanglement is that a measurement
performed on one part of the composite system determines the state of the other, whatever the
distance between them, which implies the quantum mechanical non-locality \cite{epr,bel,pan0,che}.
Quantum entanglement shared by distant objects is not only a key ingredient for the tests of the
foundations of quantum mechanics, but also a basic resource in achieving tasks of quantum
communication and quantum computation \cite{nie}. The ability to reliably create entanglement
between spatially separated objects is of particular importance for the actual implementation of
any quantum communication protocol and is also a prerequisite for distributed quantum computation.

Atomic ensembles and quantum light fields are promising candidates
for the realization of quantum computing and quantum communication
protocols \cite{she}, with long-lived atomic states constituting
quantum registers, upon which quantum logic operations can be
locally performed, and light fields providing a means of
distributing quantum information and entanglement between
different nodes in a network of registers. The workability of such
atom-light networks will depend heavily on the extent to which
propagating light fields can reliably transfer quantum states
and/or establish quantum entanglement between atoms at different
nodes of the network.

In order to make entanglement a tangible, exploitable phenomenon
one has to create entangled states of many particles, i.e.,
quantum entanglement on a macroscopic scale. So far four-ion,
four-photon, five-photon, six-atom, and eight-trapped-ion
entangled states \cite{sac,p1,p2,lei,haf} have been demonstrated
experimentally. In recent years, much attention has also been paid
to creating quantum entanglement between macroscopic atomic
samples \cite{pol,luk,dua1,jul,sor,dua2,dua3,hel,pu,dua4,cho} due
to their relatively simple experimental realization and robustness
to single particle decoherence. There are several proposals to
generate quantum entanglement between macroscopic atomic ensembles
\cite{dua3} and to explore its applications to quantum
communication \cite{dua1,dua0,jia,kuz} and quantum computation
\cite{you}. Polzik \cite{pol} proposed the first proposal to
create distant entangled macroscopic atomic ensembles through
using free propagating Einstein-Podolsky-Rosen-correlated (EPR)
light. In the proposed scheme the nonlocal correlations
transmitted by the light are mapped onto the atomic ensembles
through atoms completely absorbing the light. A similar scheme
\cite{luk} was suggested to generate entangled atomic ensembles by
trapping correlated photons in the atomic ensembles in terms of
the technique of intracavity electromagnetically induced
transparency (EIT) \cite{lu1}. Duan and coworkers \cite{dua1} then
proposed a more practical scheme to produce distant entangled
atomic ensembles in collective-spin variables by using only strong
coherent light and nonlocal Bell measurements. Shortly, a newer
method of generating entangled atomic ensembles was proposed in
Ref. \cite{dua0} through performing project measurements on the
forward-scattered Stokes light from two distant atomic ensembles.
So far the two schemes of creating entangled atomic ensembles
proposed in Refs. \cite{dua1,dua0} have been successfully
demonstrated experimentally \cite{jul,cho}. On the aspect of
atomic Bose-Einstein condensates (BECs) it has been shown that
substantial many-particle entanglement can be generated directly
in a two-component weakly interacting BEC using the inherent
interatomic interactions \cite{sor,sor1} and a spinor BEC using
spin-exchange collision interactions \cite{dua2,pu,dua4}. Based on
an effective interaction between two atoms from coherent Raman
processes, Helmerson and You \cite{hel} proposed a coherent
coupling scheme to create massive entanglement of BEC atoms. Deb
and Agarwal \cite{deb} proposed a light-Bragg-scattering scheme
for entangling two spatially separated BECs, in which a common
probe light stimulates Bragg scattering in the two BECs. The
resulting quasiparticles or particles in the two BECs get
entangled due to quantum communication between the BECs via the
probe beam. As well known, the generation of an entangled coherent
state \cite{san} is one of the most important ingredients of
quantum information processing using coherent states
\cite{jeo,enk,jeo1,ral,gla}. In a previous paper \cite{kua}, one
of the authors proposed a scheme for the generation of atom-photon
entangled coherent states in the atomic BEC which exploits EIT in
three-level atoms \cite{har}. It has been shown how to create
multistate atom-photon entangled coherent states when the
atom-photon system is initially in an uncorrelated product
coherent state. EIT is a kind of quantum interference effects
\cite{har,scu,scul,ari} and arises in three-level (or multilevel)
atomic systems. The phenomenon can be understood as a destructive
interference of the two pathways to the excited level and has been
used to demonstrate ultraslow light propagation
\cite{hau,ino,kas,bud,tur}, light storage and revivification
\cite{liu,phi,gin} in many systems including atomic BECs
\cite{hau,ino}.

In the present paper we propose a method for the generation of entangled coherent states of distant
atomic BECs via the EIT technique and optical projective measurements. Our method uses a pair of
strong coupling lasers and a pair of weak probe lasers. The coupling lasers are control lasers
which are classically treated. The two probe lasers are quantized and initially entangled. The two
pairs of lasers make the two distant BECs be in a EIT state. Under certain conditions one can
deterministically create photon-atom entangled states between two probe lasers and two distant
BECs. Then entangled states of distant BECs can be probabilistically generated through performing
projective measurements on the two outgoing probe lasers. Concretely we will propose two protocols
to create entangled coherent states of the two distant BECs corresponding to different initial
entangled states of the two probe lasers.

This paper is organized as follows. In Sec. II, we introduce the physical system involved in our
consideration, establish our model, and give an approximate analytic solution of the model. In Sec.
III, based on the analytic solution of the model we show how to create entangled states between
photons and distant atomic BECs. In Sec. IV, we show how to create entangled coherent states
between two distant atomic BECs. We shall conclude our paper with discussions and remarks in the
last section.

\section{Physical model and solution}

We consider two spatially separated BECs denoted by BEC$_1$ and
BEC$_2$, respectively. They consist of the same kind of atoms with
each atom having mass $m$. The two BECs are connected with each
other through a pair of entangled laser fields which serve as two
probe lasers. Other two coupling lasers together with the two
probe lasers make two BECs be in EIT states (Fig.1). Assume that
condensed atoms in each BEC have three internal states labelled by
$|1\rangle_i$, $|2\rangle_i$, and $|3\rangle_i$ with energies
$E_{1i}$, $E_{2i}$, and $E_{3i}$ ($i=1,2$), respectively. The two
lower states $|1\rangle_i$ and $|3\rangle_i$ are Raman coupled to
the upper state  $|2\rangle_i$ via, respectively, a quantized
probe-laser field and a classical coupling-laser fields of
frequencies $\omega_{i1}$ and $\omega_{i2}$ in the $\Lambda$-type
configuration indicated in Fig. 2. The interaction scheme is shown
in Fig. 1. The condensed atoms in these internal states are
subject to isotropic harmonic trapping potentials $V_{in}({\bf
r})$ for $i=1,2,3$ and $n=1,2$, respectively. Furthermore, the
atoms in the two BECs interact with each other via elastic
two-body collisions with the $\delta$-function potentials
$V^{n}_{ij}({\bf r}-{\bf r}')=U^{n}_{ij}\delta ({\bf r}-{\bf r}')$
where $U^{n}_{ij}=4\pi\hbar^2a^{n}_{ij}/m$ with $m$ and
$a^{n}_{ij}$, respectively, being the atomic mass and  the
$s$-wave scattering length. A good experimental candidate of this
system is the sodium atom condensate for which there exist
appropriate atomic internal levels and external laser fields to
form the $\Lambda$-type configuration which has been used to
demonstrate ultraslow light propagation \cite{hau} and
amplification of light and atoms \cite{ino1} in atomic BECs.

The second quantized Hamiltonian to describe the system at zero
temperature is given by
\begin{equation}
\label{1}
\hat{H}=\sum^{2}_{n=1}\left(\hat{H}_{Pn}+\hat{H}_{An}+\hat{H}_{In}+\hat{H}_{Cn}\right),
\end{equation}
where $\hat{H}_{An}$ gives the free evolution of the atomic
fields, $\hat{H}_{In}$ describes the dipole interactions between
the atomic fields and laser fields for each BEC, and
$\hat{H}_{Cn}$ represents interatomic two-body collision
interactions in each BEC.

\begin{figure}[htb]
\begin{center}
\includegraphics[width=8cm,height=5cm]{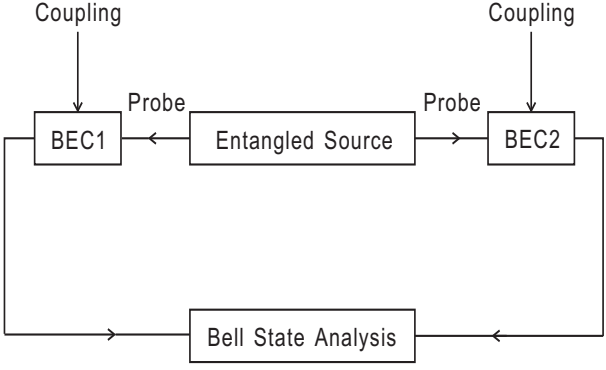}
\end{center}
\vskip 0.2cm \caption{The schematic diagram of creating entangled
coherent states for two distant BECs via EIT. Two pairs of
coupling and probe lasers are applied on two BECs, respectively.
The entangled source produces two entangled probe lasers. After
leaving two BECs, the two outgoing probe lasers enter Bell state
analyzer to perform projective measurements.} \label{fig1}
\end{figure}

The free evolution of the two quantized probe laser fields is
governed by the Hamiltonian
\begin{equation}
\label{2}
\hat{H}_{Pn}=\sum^{2}_{n=1}\left(\hbar\omega_n\hat{a}^{\dagger}_n\hat{a}_n\right),
\end{equation}
where $\omega_n$ is the frequency of the $n$-th probe laser, and $
\hat{a}^{\dagger}_n$ and $\hat{a}_n$ are the corresponding photon
creation and annihilation operators for the $n$-th probe laser
field, satisfying the boson communication relation $[\hat{a}_n,
\hat{a}^{\dagger}_n]=1$.

The free atomic Hamiltonian is given by
\begin{eqnarray}
\label{3} \hat{H}_{An}&=&\sum^{3}_{i=0}\int d{\bf x}
\hat{\psi}^{\dagger}_{i n}({\bf x}) \hat{H}_{in}
\hat{\psi}_{i n}({\bf x}), \nonumber\\
\hat{H}_{in}&=&-\frac{\hbar^2}{2m}\nabla^2 +V_{i n}({\bf x})+E_{i
n},
\end{eqnarray}
where $E_{i n}$ are internal energies for the three internal
states, $\hat{\psi}_{i n}({\bf x})$ and $\hat{\psi}^{\dagger}_{i
n} ({\bf x})$ are the boson  annihilation and creation operators
for the $|i\rangle$-state atoms at position ${\bf x}$ for the
$n$-th BEC, respectively, they satisfy the standard boson
commutation relation $[\hat{\psi}_{i n}({\bf x}),
\hat{\psi}^{\dagger}_{j m}({\bf x}')]=\delta_{ij}\delta_{n m
}\delta({\bf x}-{\bf x}')$ and $[\hat{\psi}_{i n}({\bf x}),
\hat{\psi}_{j m}({\bf x}')]=0$, and $[\hat{\psi}^{\dagger}_{i
n}({\bf x}), \hat{\psi}^{\dagger}_{j m}({\bf x}')]=0$.

The atom-laser interactions in the dipole approximation can be
described by the following Hamiltonian
\begin{eqnarray}
\label{4}
\hat{H}_{In}&=&-\hbar \int d{\bf x}  \left [g_n\hat{a}_n\hat{\psi}^{\dagger}_{2n}({\bf x})\hat{\psi}_{1n}({\bf x})e^{i({\bf k_1}\cdot {\bf x}-\omega_1t)} \right. \nonumber \\
& &\left.+\frac{1}{2}\Omega_n\hat{\psi}^{\dagger}_{2n}({\bf x})
\hat{\psi}_{3n}({\bf x})e^{i({\bf k_2}\cdot {\bf
x}-\omega_2t)}+\tt{H.c.}\right ],
\end{eqnarray}
where dipole coupling constants are defined by $g_n=\mu_{21}{\cal
E}_{1n}/\hbar$ and $\Omega_n=\mu_{23}E_{2n}/\hbar$ with $\mu_{ij}$
denoting a transition dipole-matrix element between states
$|i\rangle$ and $|j\rangle$, ${\cal
E}_{1n}=\sqrt{\hbar\omega_{1n}/2\epsilon_0V}$ being the electric
field per photon for the quantized probe light of frequency
$\omega_{1n}$ in a mode volume $V$, and $E_{2n}$ being the
amplitude of the electric field for the classical coupling light
of frequency $\omega_2$, ${\bf k}_{1n}$ and ${\bf k}_{2n}$ are
wave vectors of correspondent laser fields.

The interatomic collision Hamiltonian is taken to be the following
form
\begin{eqnarray}
\label{5} \hat{H}_{Cn}&=&\frac{2\pi\hbar^2}{m} \int d{\bf x}\left
[\sum^3_{i=1}a^{sc}_{in}\hat{\psi}^{\dagger}_{in}({\bf x})
\hat{\psi}^{\dagger}_{in}({\bf x})\hat{\psi}_{in}({\bf
x})\hat{\psi}_{in}({\bf x})
\right. \nonumber \\
& &\left.+\sum_{i\neq j}2a^{sc}_{ij}
\hat{\psi}^{\dagger}_{in}({\bf x}) \hat{\psi}^{\dagger}_{jn}({\bf
x})\hat{\psi}_{in}({\bf x})\hat{\psi}_{jn}({\bf x}) \right ],
\end{eqnarray}
where $a^{sc}_i$ is  the $s$-wave scattering length of condensed
atoms  in the internal state $|i\rangle$ and $a^{sc}_{ij}$ that
between condensed atoms in the internal states $|i\rangle$ and
$|j\rangle$.

\begin{figure}[htb]
\begin{center}
\includegraphics[width=7cm]{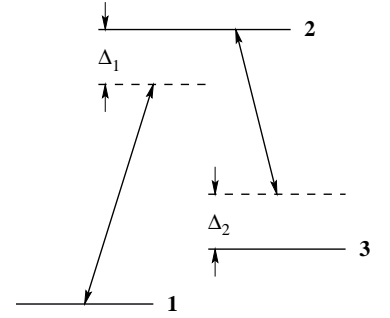}
\end{center}
\vskip 0.2cm \caption{Three-level $\Lambda$-type atoms coupled to
two laser fields with the detunings $\Delta_1$ and $\Delta_2$,
respectively.} \label{fig2}
\end{figure}

For a weakly interacting BEC at zero temperature there are no
thermally  excited atoms and the quantum depletion is negligible,
the motional state is frozen to be approximately the ground state.
One may neglect all modes except for the condensate mode and
approximately factorize the atomic field operators as
$\hat{\psi}_{in}({\bf x})\approx \hat{b}_{in}\phi_{in}({\bf x})$
where $\phi_{in}({\bf x})$  is a normalized wavefunction for the
atoms in the BEC in the internal state $|i\rangle$,  which is
given by the ground state of the following Schr\"{o}dinger
equation
\begin{equation}
\label{6} \left [-\frac{\hbar^2}{2m}\nabla^2 +V_{in}({\bf
x})+E_{in}\right ] {\phi}_{in}({\bf
x})=\hbar\nu_{in}{\phi}_{in}({\bf x}),
\end{equation}
where $\hbar\nu_{in}$ is the energy of the mode $i$, and
$\nu_{in}$ denotes the frequency to the free evolution of the
condensate in the internal state $|i\rangle$.

Substituting the single-mode expansions of the atomic field
operators into Eqs. (\ref{3}-\ref{5}), we arrive at the following
approximate Hamiltonian
\begin{equation}
\label{7}\hat{H}=\hat{H}_1 + \hat{H}_2,
\end{equation}
where the two commutable parts of the Hamiltonian are given by
\begin{eqnarray}
\label{8}
\hat{H}_n&=&\hbar\omega_{n}\hat{a}^{\dagger}_n\hat{a}_n + \hbar\sum^3_{i=1}\nu_{in}\hat{b}^{\dagger}_{in}\hat{b}_{in} \nonumber \\
&& -\hbar\left
[g_{1n}\hat{a}_n\hat{b}^{\dagger}_{2n}\hat{b}_{1n}e^{-i\omega_1t}
+g_{2n}\hat{b}^{\dagger}_{2n}\hat{b}_{3n}e^{-i\omega_2t}+
\tt{H.c.} \right ]
\nonumber \\
&&+ \hbar\sum^3_{i=1}\lambda_{in}\hat{b}^{\dagger
2}_{in}\hat{b}^2_{in} + \hbar\sum_{i\neq
j}\lambda_{ijn}\hat{b}^{\dagger}_{in}\hat{b}_{in}
\hat{b}^{\dagger}_{jn}\hat{b}_{jn},
\end{eqnarray}
where the modes $\hat{a}_n$ ($n=1,2$) correspond to the two probe laser fields, the modes
$\hat{b}_{in}$ ($n=1,2,$ and $i=1,2,3$) to atomic fields in the three internal state, respectively.
Here $g_{1n}$ and $g_{2n}$ denote the laser-atom dipole interactions. They are defined by
\begin{eqnarray}
\label{9} g_{1n}&=&g_n\int d{\bf x} \phi^*_{2n}({\bf
x})\phi_{1n}({\bf
x})e^{i{\bf k}_1.{\bf   x}}, \nonumber \\
g_{2n}&=&\frac{1}{2}\Omega_n\int d{\bf x} \phi^*_{2n}({\bf
x})\phi_{3n}({\bf x})e^{i{\bf k}_2.{\bf x}}.
\end{eqnarray}
And $\lambda_{in}$ and $\lambda_{ijn}$ ($i, j=1, 2, 3$) describe interatomic interactions ($s$-wave
elastic scattering) given by
\begin{eqnarray}
\label{10} \lambda_{ijn}&=&\frac{4\pi\hbar^2a^{sc}_{ij}}{m}\int
d{\bf x} |\phi_{in}({\bf x})|^2|\phi_{jn}({\bf x})|^2, \hspace{0.3cm}(i\neq j),  \nonumber \\
\lambda_{in}&=& \frac{2\pi\hbar^2a^{sc}_i}{m}\int d{\bf
x}|\phi_{in}({\bf x})|^4.
\end{eqnarray}

Properly choosing a unitary transformation, we can transfer the
time-dependent Hamiltonian (\ref{7}) to the following
time-independent Hamiltonian
\begin{equation}
\label{11}  \hat{H}_{I}=\hat{H}_{I1}+ \hat{H}_{I2},
\end{equation}
where $\hat{H}_{In}$ ($n=1, 2$) can take the following form
\begin{eqnarray}
\label{12}
 \hat{H}_{In} &=&\hbar(\Delta_{1n}-\Delta_{2n})\hat{b}^{\dagger}_{3n}\hat{b}_{3n}+\Delta_{1n}\hat{b}^{\dagger}_{2n}\hat{b}_{2n}
\nonumber \\
& &-\hbar[g_{1n}\hat{a}_1\hat{b}^{\dagger}_{2n}\hat{b}_{1n}+
g_{2n}\hat{b}^{\dagger}_{2n}\hat{b}_{3n}+ \tt{H.c.}]
\nonumber \\
& & + \hbar\sum^3_{i=1}\lambda_{in}\hat{b}^{\dagger
2}_{in}\hat{b}^2_{in} + \hbar\sum_{i\neq
j}\lambda_{ijn}\hat{b}^{\dagger}_{in}\hat{b}_{in}
\hat{b}^{\dagger}_{jn}\hat{b}_{jn},
\end{eqnarray}
where $\Delta_{1n}=\nu_{2n}-\nu_{1n}-\omega_{1n}$ and
$\Delta_{2n}=\nu_{2n}-\nu_{3n}-\omega_{2n}$ are the detunings of
the corresponding probe and coupling laser beams, respectively.

We consider the situation of the ideal EIT which is attained only when the system is at the
two-photon resonance with the detuning $\Delta =\Delta_{1n}=\Delta_{2n}$ and the involved lasers
are applied adiabatically. Initially the lasers are outside the BEC medium in which all atoms are
in their ground state, i.e., the internal state $|1\rangle_i$ ($i=1,2$). These condensed atoms are
generally in a superposition state of the state $|1\rangle_i$ and the state $|3\rangle_i$  when
they are in EIT. However, when the coupling laser is much stronger than the probe laser, atomic
population at the intermediate level approaches zero while the upper level is unpopulated when EIT
occurs \cite{scul}. Hence, when the BECs are in EIT states, under the condition of
$(g_{1n}/g_{2n})^{2}<<1$ the terms which involve $\hat{b}^{\dagger}_{2n}\hat{b}_{2n}$ and
$\hat{b}^{\dagger}_{3n}\hat{b}_{3n}$ in the Hamiltonian (\ref{8}) may be neglected, and from the
Hamiltonian (\ref{8}) we can adiabatically eliminate the atomic field operators in the internal
states $|2\rangle$ and $|3\rangle$ \cite{kua}. Then, the resulting effective Hamiltonian contains
only the atomic field operators in internal states $|1\rangle_n$ and the probe field operators and
have the following simple form
\begin{equation}
\label{13}
\hat{H}_{eff}=\sum^2_{n=1}\left(\omega'_{n}\hat{b}^{\dagger}_{n}\hat{b}_{n}
+2\omega'_{n}\hat{b}^{\dagger}_{n}\hat{b}_{n}\hat{a}^{\dagger}_n\hat{a}_n+\lambda_n
\hat{b}^{\dagger 2}_{n}\hat{b}^2_{n}\right),
\end{equation}
where we have set $\hat{b}_{n}=\hat{b}_{1n}$,
$\hat{b}^{\dagger}_{n}=\hat{b}^{\dagger}_{1n}$  and
$\lambda_n=\lambda_{1n}$ for the simplification of notations, and
we have also taken $\hbar=1$ and introduced
\begin{equation}
\label{14} \omega'_{n}=-\frac{2|g_{1n}|^2}{\Delta_{1n}}.
\end{equation}

Obviously, the effective four-mode Hamiltonian (\ref{13}) is
diagonal in the Fock space with the bases defined by
\begin{equation}
\label{15} |n_1, n_2,m_1, m_2\rangle =\frac{\hat{a}^{\dagger
n_1}_1\hat{a}^{\dagger n_2}_2\hat{b}^{\dagger
m_1}_1\hat{b}^{\dagger
m_2}_2}{\sqrt{n_1!n_2!m_1!m_2!}}|0,0,0,0\rangle ,
\end{equation}
which are eigenstates of the effective Hamiltonian with the
eigenvalues given by the following expression
\begin{equation}
\label{16}
 E_{n_1,n_2,m_1,m_2}=\sum^2_{i=1}\left[\omega'_i m_i + 2\omega'_in_im_i + \lambda_im_i(m_i-1)\right],
\end{equation}
where $n_i$ and $m_i$ take non-negative integers.

It should be mentioned that in most textbooks on EIT
\cite{scul,ari}, both the probe and coupling lasers are treated
classically, the decay of excited states is included in the
dynamics of the internal states through a set of density-matrix
equations of the atomic system,  the decay promotes the trapping
of the atom into a dark state. Contrary to the usual treatment of
EIT, the probe laser is quantized in present paper. A limitation
of our present treatment is that we have ignored the decay rates
of various levels. However, this ignorance of the decay rates is a
good approximation for the adiabatic EIT that we study in the
present paper, since the adiabatic  EIT is insensitive to any
possible decay of the top level \cite{lm}.

\section{Generation of entangled state between photons and distant BECs}

In this section, as one of the key ingredients to create entangled
coherent states of the two distant BECs we investigate the
generation of atom-photon entangled state between the two probe
laser fields and two distant BEC system in terms of analytic
solution obtained in the previous section. Since there does not
exist direct interaction between the two distant BECs under our
consideration, entangled states between them can not be created
through interaction between two distant BECs if they are initially
not entangled. However, if the two probe laser fields are
initially in an entangled state, it is possible to create
entangled state of the two distant BECs through performing
appropriate quantum measurements upon output state of the two
probe laser fields. In some sense we can say that entanglement is
transferred from the two probe laser fields to the two distant
BECs. An initial entangled state of the two probe laser fields can
be regarded as a quantum channel to create quantum entanglement
between the two distant BECs. Once the quantum channel is given,
one of important steps of creating entangled distant BECs is to
generate photon-atom entangled state. In order to be specific, we
here use an entangled single-photon channel to show how to create
a hybrid entangled state between the two probe laser fields and
two distant BECs to make preparation for obtaining entangled
states of distant BECs which will be discussed in the next
section.

We assume that the two distant BECs are initially uncorrelated and
they are in a two-mode product coherent state $|\alpha_1,
\alpha_2\rangle\equiv |\alpha_1\rangle
\otimes|\alpha_2\rangle=D_{\hat{b}_1}(\alpha_1)D_{\hat{b}_2}(\alpha_2)|0,0\rangle
$ with the displacement operators being defined by
$D_{\hat{b}_i}(\alpha_i)=\exp(\alpha_i\hat{b}^{\dagger}_i
-\alpha^*_i\hat{b}_i)$, while the entangling source produces a
pair of the probe lasers which is in the entangled single-photon
state
\begin{equation}
\label{17} |{\tt B}(\theta, \varphi)\rangle =\cos\theta|0,0\rangle
+\sin\theta e^{i\varphi}|1,1\rangle.
\end{equation}

Then the joint initial state of the atom-photon system is given by
\begin{equation}
\label{18} |\Phi(0)\rangle =|{\tt B}(\theta,
\varphi)\rangle\otimes|\alpha_1, \alpha_2\rangle.
\end{equation}

Assume that the lasers are adiabatically applied at the time of
$t=0$, then from Eqs. (\ref{13}-\ref{18}) it is straightforward to
show that after an interaction time $t$ the joint state of the
system will be given by
\begin{eqnarray}
\label{19} |\Phi(t)\rangle
&=&\cos\theta|0,0\rangle\otimes|\phi_{\alpha_1 \alpha_2}(t)\rangle
\nonumber\\
& &+ \sin\theta e^{i\varphi}|1,1\rangle\otimes|\phi'_{\alpha_1
\alpha_2}(t)\rangle,
\end{eqnarray}
where we have introduced two  wave functions for the two distant
BECs
\begin{eqnarray}
\label{20} |\phi_{\alpha_1 \alpha_2}(t)\rangle
&=&|\phi_{\alpha_1}(t)\rangle\otimes|\phi_{\alpha_2}(t)\rangle,
\nonumber \\
|\phi'_{\alpha_1 \alpha_2}(t)\rangle
&=&|\phi'_{\alpha_1}(t)\rangle\otimes|\phi'_{\alpha_2}(t)\rangle,
\end{eqnarray}
where the two wave functions of the single BECs are defined by
\begin{eqnarray}
\label{21} |\phi_{\alpha_i}(t)\rangle
&=&\exp\left\{-it[\omega'_i\hat{n}_{b_i}+\lambda_i\hat{n}_{b_i}(\hat{n}_{b_i}-1)]\right\}|\alpha_i\rangle,
\nonumber \\
|\phi'_{\alpha_i}(t)\rangle
&=&\exp\left\{-it[3\omega'_i\hat{n}_{b_i}+\lambda_i\hat{n}_{b_i}(\hat{n}_{b_i}-1)]\right\}|\alpha_i\rangle.
\end{eqnarray}

In the following we shall see that starting with the state given
by Eq. (\ref{19}), different atom-photon entangled states can be
obtained at different evolution times of the joint system under
our consideration through adjusting various interaction strengths
and the detunings. In order to see this, for simplicity we suppose
that $g_{1n}=g_{2n}=g$, $\omega'_1=\omega'_2=\omega'$,
$\lambda_{1}=\lambda_{2}=\lambda$, and introduce a scaled time
$\tau=\lambda t$, then the states of the single BECs given by Eq.
(\ref{20}) become the following generalized coherent states
\cite{tit,bia,sto}
\begin{eqnarray}
\label{22} |\phi_{\alpha_i}(\tau)\rangle
&=&e^{-|\alpha_i|^2/2}\sum^{\infty}_{m_i=0}e^{i\tau
\theta_{m_i}}\frac{\alpha^{m_i}_i}{\sqrt{m_i!}}|m_i\rangle,
\nonumber \\
|\phi'_{\alpha_i}(\tau)\rangle
&=&e^{-|\alpha_i|^2/2}\sum^{\infty}_{m_i=0}e^{i\tau
\theta'_{m_i}}\frac{\alpha^{m_i}_i}{\sqrt{m_i!}}|m_i\rangle,
\end{eqnarray}
where two running frequencies $\theta_{m_i}$ and $\theta'_{m_i}$
are defined by
\begin{equation}
\label{23} \theta_{m_i}=(K+1)m_i-m^2_i,
\hspace{0.3cm}\theta'_{m_i}=(3K+1)m_i-m^2_i,
\end{equation}
where we have introduced a real effective interaction parameter
$K$ defined by
\begin{equation}
\label{24} K=-\frac{2|g|^2}{\lambda \Delta},
\end{equation}
which describes an effective interaction induced by three
adjustable parameters of the involved lasers and two distant BECs:
the dipole interaction strength $g$, the interatomic interaction
strength $\lambda$, and the two-photon detuning $\Delta$. The
effective interaction parameter $K$ can takes positive or negative
values which depends on the signs of $\lambda$ and $\Delta$. For
an atomic BEC with a positive $s$-wave scattering length, $K$ is
positive (negative) when the two-photon detuning $\Delta$ is
positive (negative). These indicate that there is a large space to
adjust values of the effective interaction parameter $K$. From
Eqs. (\ref{19}) to (\ref{23}) we can see that the time evolution
characteristic of the system under our consideration is completely
determined by the effective interaction parameter.

These generalized coherent states given by  Eq. (\ref{22}) differ
from a conventional Glauber coherent state by an extra phase
factor appearing in the decomposition of such states into a
superposition of Fock states. They can always be represented as a
continuous sum (integral) form of Glauber coherent states. And
under appropriate periodic conditions, they can reduce to discrete
superpositions of coherent states. In order to see this, we
express states given by Eq. (\ref{22}) as the following integral
form
\begin{eqnarray}
\label{25} |\phi_{\alpha_i}(\tau)\rangle
&=&\frac{1}{2\pi}\int^{2\pi}_0 d\phi_i
 f(\phi_i,\theta_{m_i})\left|\alpha_ie^{i\phi_i}\right\rangle, \nonumber \\
|\phi'_{\alpha_i}(\tau)\rangle &=&\frac{1}{2\pi} \int^{2\pi}_0
d\phi_i
 f(\phi_i,\theta'_{m_i})\left|\alpha_ie^{i\phi_i}\right\rangle,
\end{eqnarray}
where the two phase functions are introduced with the following
expressions
\begin{eqnarray}
\label{26}
 f(\phi_i,\theta_{m_i})&=&\exp\left[i\left(\tau\theta_{m_i} - m_i\phi_i
\right)\right],\nonumber \\
 f(\phi_i,\theta'_{m_i})&=&\exp\left[i\left(\tau\theta'_{m_i} - m_i\phi_i
\right)\right].
\end{eqnarray}

It should be kept in mind that we want to create entangled
coherent states for two distant BECs what we expect.  Eq.
(\ref{25}) indicates that both the state $|\phi_{\alpha_1
\alpha_2}(\tau)\rangle $ and the state $|\phi'_{\alpha_1
\alpha_2}(\tau)\rangle $ are  continuous superposition states of
two-mode product coherent states. From Eqs. (\ref{22}-\ref{26}) we
can see that the values of the effective interaction parameter $K$
may seriously affect the form of the states given by  Eq.
(\ref{25}). Of particular interesting is the situation where $K$
takes values of nonzero integers. In this case, both
$\theta_{m_i}$ and $\theta'_{m_i}$ take integers. Making use of
Eq. (\ref{22}) and Eq. (\ref{23}), from Eqs.  (\ref{19}) and
(\ref{20}) we can see that
$|\Phi(\tau+2\pi)\rangle=|\Phi(\tau)\rangle$, which implies that
the time evolution of the joint wavefunction given by Eq.
(\ref{19}) is a periodic evolution with respect to the scaled time
$\tau$ with the period $2\pi$. On the other hand, suppose that the
scaled interaction time $\tau$ takes its values in the following
manner
\begin{equation}
\label{27} \tau= \frac{M}{N}2\pi,
\end{equation}
where $M$ and $N$ are mutually prime integers, then we can find
that
\begin{eqnarray}
\label{28}  f(\phi_i,\theta_{m_i})&=&f(\phi_i,\theta_{m_i+N}),
\nonumber\\
f(\phi_i,\theta'_{m_i})&=&f(\phi_i,\theta'_{m_i+N}),
\end{eqnarray}
which means that the two exponential phase functions in the states
given by Eq. (\ref{25}) are also periodic functions with respect
to $m_i$ with the same period $N$. If $\tau$ takes its values
according to Eq. (\ref{27}), as a fraction of the period, then the
two states given by Eq. (\ref{25}) become  discrete superposition
states of product coherent states, and they can be expressed as
follows
\begin{eqnarray}
\label{29} \left|\phi_{\alpha_1 \alpha_2}\left(\tau=2\pi
M/N\right)\right\rangle&=& \prod^2_{i=1} \sum^{N}_{r_i=1}
c_{r_i}\left|\alpha_i
e^{i\varphi_{r_i}}\right\rangle, \nonumber \\
\left|\phi'_{\alpha_1 \alpha_2}\left(\tau=2\pi
M/N\right)\right\rangle&=& \prod^2_{i=1} \sum^{N}_{r_i=1}
c'_{r_i}\left|\alpha_i e^{i\varphi_{r_i}}\right\rangle,
\end{eqnarray}
the running phase is defined by the following expression
\begin{equation}
\label{30} \varphi_{r_i}=\frac{2\pi}{N}r_i, \hspace{0.5cm}
(r_i,=1,2, \cdots, N),
\end{equation}
and different superposition coefficients are given by
\begin{eqnarray}
\label{31} c_{r_i}&=&\frac{1}{N}\sum^{N}_{m_i=1}
\exp\left\{-\frac{2\pi i }{N}\left[m_ir_i -M
\theta_{m_i}\right]\right\}, \nonumber \\
c'_{r_i}&=&\frac{1}{N}\sum^{N}_{m_i=1} \exp\left\{-\frac{2\pi i
}{N}\left[m_ir_i -M \theta'_{m_i}\right]\right\}.
\end{eqnarray}

Now it is easy to see that after an interaction time $\tau=2\pi
M/N$ with the lasers the joint state given by Eq. (\ref{19})
becomes
\begin{widetext}
\begin{eqnarray}
\label{32} \left|\Phi\left(\tau =\frac{M}{N}2\pi
\right)\right\rangle
&=&\cos\theta|0,0\rangle\otimes\sum^{N}_{r_1,r_2=1}c_{r_1}c_{r_2}|\alpha_1
e^{i\varphi_{r_1}}, \alpha_2 e^{i\varphi_{r_2}} \rangle
\nonumber\\
&&+ \sin\theta
e^{i\varphi}|1,1\rangle\otimes\sum^{N}_{r_1,r_2=1}c'_{r_1}c'_{r_2}|\alpha_1
e^{i\varphi_{r_1}}, \alpha_2 e^{i\varphi_{r_2}} \rangle.
\end{eqnarray}
\end{widetext}

The state given by Eq. (\ref{32}) is generally an atom-photon
entangled state between the two probe laser fields and the two
distant BECs. It is interesting to note that it is a hybrid
entangled state in which one subsystem (photon) is discrete while
another (BEC) is continuous, since it exhibits quantum
entanglement between two distinct degrees of freedom
\cite{mon,bru,chen,hal}. In principle, hybrid entangled state can
serve as a valuable resource in quantum information processing and
build up a bridge between quantum information protocols of
discrete and continuous variables. As a specific example of
creating atom-photon hybrid entangled states, we discuss the case
of $K=1$. Assume that the probe lasers and the condensates are
initially in the state given by (\ref{18}) with $\theta=\pi/4$ and
$\varphi=0$. When $K=1$, $N=4$, and $M=1$, from Eqs. (\ref{31})
and (\ref{32}) we find that after the interaction times
$\tau=\pi/2$ the joint wave function given by Eq. (\ref{32})
becomes the following hybrid entangled state
\begin{eqnarray}
\label{33} |\Phi(\tau=\pi/2)\rangle&=&\mathcal{N}\left[|B (\pi/4,
0)\rangle\otimes |E(\alpha_1,-\alpha_2)\rangle \right.
\nonumber \\
&& \left.+ i|B (-\pi/4, 0)\rangle\otimes
|E(\alpha_1,\alpha_2)\rangle\right],
\end{eqnarray}
where  $|B (\pm\pi/4, 0)\rangle$ are two optical Bell states
defined by
\begin{equation}
\label{34} |B (\pm\pi/4, 0)\rangle =\frac{1}{\sqrt2}(|0,0\rangle
\pm |1,1\rangle),
\end{equation}
while  $|E(\alpha_1,\pm\alpha_2)\rangle$  are two normalized
entangled coherent states \cite{hir} for the two BECs, they are
defined by
\begin{equation}
\label{35}  |E(\alpha, \beta)\rangle=\frac{1}{\sqrt{N_{\alpha
\beta}}}( |\alpha, \beta\rangle + |-\alpha,-\beta\rangle),
\end{equation}
where the normalized constant is given by
\begin{equation}
\label{36} N_{\alpha
\beta}=2\left\{1+\exp[-2(|\alpha|^2+|\beta|^2)]\right\},
\end{equation}
which gives the normalized constant in Eq. (\ref{33}) by the form
$\mathcal{N}= (N_{\alpha \beta}/2)^{1/2}$. When $\beta=\pm\alpha$,
the two states $|E(\alpha_1,\pm\alpha_2)\rangle$ become quasi-Bell
states which will defined in the next section.

Atom-photon entangled states given by Eqs. (\ref{32}) and
(\ref{33}) are important results of the present paper. They
indicate that we can deterministically generate atom-photon
entangled states between the two probe lasers and the two distant
BECs. In particular, we note that the entangled state given by Eq.
(\ref{33}) is of three interesting characters. First of all, the
state given by Eq. (\ref{33}) is a hybrid entangled state between
atoms and photons where atoms and photons are entangled, photons
are well suited for transmission over long distances and carry
away information on the state of atoms while atoms have the
advantage of being more easily stored. Such an entangled state
provides a possibility for a communication link in a quantum
network. Secondly, the state given by Eq. (\ref{33}) is a type of
entangled states between two pairs of entangled states where one
entangled pair is Bell states of photons while the other is
entangled coherent states of atoms. In particular, when
$\alpha_2=\pm\alpha_1$,it is an entangled state between a pair of
optical Bell states and a pair of atomic quasi-Bell states. The
practical development of such kind of entangled states is of
importance for exploring quantum nonlocality and testing Bell's
inequality. Thirdly, the state given by Eq. (\ref{33}) is also a
hybrid entangled state between discrete-variable states and
continuous variable states, such a kind of entangled states may
open new ways to carry out quantum information processing. In what
follows we will observe that these atom-photon entangled states in
this section produced give a supply for the generation of
entangled coherent states of the two distant BECs which will be
investigated in the next section.

\section{Generation of entangled coherent states for distant BECs}

From the discussions of the previous section we can see that
different atom-photon entangled states between the probe lasers
and the two distant BECs can be obtained using different initial
entangled states (quantum channels) of the probe lasers. In this
section we study the generation of entangled coherent states of
the two distant BECs using different quantum channels. Concretely,
we will propose two schemes to create entangled coherent states of
the two distant BECs. One is a single-photon scheme, the other is
a multiphoton scheme. An entangled single-photon state is used to
serve the quantum channel in the single-photon protocol while an
entangled coherent-state acts as the quantum channel in the
multiphoton scheme.

\begin{center}
{\bf A. The entangled single-photon-state scheme}
\end{center}

In this subsection we discuss how to produce entangled coherent
states between two distant BECs using the single-photon quantum
channel. In this case, the two probe lasers are initially in the
entangled single-photon state  $ |B (\theta, \varphi)\rangle $
given by Eq. (\ref{17}) in which there is an one-photon state in
each probe mode while the two BECs are initially in a product
coherent state. Then the joint initial state of the atom-photon
system is given by (\ref{18}). We note that at a time $t$ the
state of the system under our consideration is given by
(\ref{19}), and it can be expressed in terms of optical Bell bases
of probe photons as the following form
\begin{eqnarray}
\label{37} |\Phi(t)\rangle &=&\frac{1}{\sqrt2}[|B
(\pi/4,0)\rangle\otimes|\psi_{\alpha_1 \alpha_2}(t)\rangle_+
\nonumber\\
&&+ |B (-\pi/4,0)\rangle\otimes|\psi_{\alpha_1
\alpha_2}(t)\rangle_-],
\end{eqnarray}
where  $|B (\pm\pi/4, 0)\rangle$ are two optical Bell states defined by Eq. (\ref{34}) and we
introduce the following superposition states of the two BECs
\begin{equation}
\label{38} |\psi_{\alpha_1
\alpha_2}(t)\rangle_{\pm}=\cos\theta|\phi_{\alpha_1
\alpha_2}(t)\rangle \pm \sin\theta e^{i\varphi}|\phi'_{\alpha_1
\alpha_2}(t)\rangle,
\end{equation}
where states $|\phi_{\alpha_1 \alpha_2}(t)\rangle$ and $|\phi'_{\alpha_1 \alpha_2}(t)\rangle$ are
defined in Eq. (\ref{20}), they are not entangled states during the time evolution of the whole
system under our consideration.

Starting with the state given by Eq. (\ref{37}), we can see that the superposition states of the
two distant BECs defined by Eq. (\ref{38}) can be probabilistically produced through performing
optical Bell measurements upon the outgoing states of the two probe lasers. From the discussions of
the previous section we can see that both $|\phi_{\alpha_1 \alpha_2}(t)\rangle$ and
$|\phi'_{\alpha_1 \alpha_2}(t)\rangle$ are generally superposed coherent states. Hence, in general
the states $|\psi_{\alpha_1 \alpha_2}(t)\rangle_{+}$ and $|\psi_{\alpha_1 \alpha_2}(t)\rangle_{-}$
are entangled coherent states. In particular, when $\tau=2\pi M/N$ with $M$ and $N$ being prime
numbers, we have the following entangled coherent states
\begin{equation}
\label{39} |\psi_{\alpha_1 \alpha_2}(\tau=2\pi M/N)\rangle_{\pm}=\sum^N_{r_1,r_2=1}C^{\pm}_{r_1
r_2}|\alpha_1e^{i\varphi_{r_1}}\rangle |\alpha_2e^{i\varphi_{r_2}}\rangle ,
\end{equation}
where the superposed coefficients are given by
\begin{equation}
\label{40} C^{\pm}_{r_1 r_2}=\cos\theta c_{r_1}c_{r_2}\pm
\sin\theta e^{i\varphi}c'_{r_1}c'_{r_2}.
\end{equation}

It is interesting to note that the entangled coherent states given
by Eq. (\ref{39}) are higher-dimensional entangled coherent states
which are superposed states of $N^2$ linearly independent product
coherent states $|\alpha_1e^{i\varphi_{r_1}}\rangle
|\alpha_2e^{i\varphi_{r_2}}\rangle$ with $r_i=1,2,\cdots, N$ and
$i=1,2$. However, from Eqs. (\ref{38}-\ref{40}) we can also see
that when the two probe lasers are initially not entangled, i.e.,
$\theta=k\pi$ or $\theta=(2k+1)\pi/2$ with $k$ being an arbitrary
integer, the states given by Eqs. (\ref{38}) and (\ref{39}) are
not entangled states. This means that only when the two probe
lasers are initially entangled, it is possible to generate
entangled states of the two distant BECs through performing
optical Bell measurements. In this sense we can say that quantum
entanglement between the two probe lasers is transferred to the
two distant BECs through performing optical Bell measurements.

In order to be more specific, let us consider the case of $K=1$,
$N=4$, and $M=1$. From Eq. (\ref{19}) and Eq. (\ref{29}) we find
the atom-photon state is given by Eq. (\ref{30}). Substituting Eq.
(\ref{29}) into Eq. (\ref{38})  we arrive at the following
entangled coherent states between the two distant BECs
\begin{eqnarray}
\label{41} |\psi_{\alpha_1
\alpha_2}(\tau=\pi/2)\rangle_{\pm}&=&C_{\pm}(\theta,\varphi)||E(\alpha_1,
-\alpha_2)\rangle \nonumber
\\ && + iC_{\mp}(\theta,\varphi)||E(\alpha_1, \alpha_2)\rangle,
\end{eqnarray}
where $C_{\pm}(\theta,\varphi)=\cos\theta\pm \sin\theta
e^{i\varphi}$, and $||E(\alpha_1, \pm\alpha_2)\rangle\equiv
|\alpha_1, \pm\alpha_2\rangle + |\pm\alpha_1, \alpha_2\rangle$ are
two un-normalized entangled coherent states.

Especially, the state given  by Eq. (\ref{41}) indicates that for the case of $K=1$  if initially
the two probe lasers is in an optical Bell state given by Eq. (\ref{17}) with $\theta=\pi/4$ and
$\varphi=0$, then after an interaction time $\tau=\pi/2$ optical Bell measurements upon the two
outgoing probe lasers can induce the following two-state entangled coherent states between the two
distant BECs (after normalization)
\begin{eqnarray}
\label{42} \left|\psi_{\alpha_1
\alpha_2}(\tau=\pi/2)\right\rangle_{\pm}&=&|E(\alpha_1,\mp\alpha_2)\rangle,
\end{eqnarray}
where $|E(\alpha_1,\pm\alpha_2)\rangle$ are entangled coherent
states defined in Eq. (\ref{35}). They are quantum superpositions
of four macroscopically distinguishable and linearly independent
coherent states $|\alpha_1, \alpha_2 \rangle$, $|\alpha_1,
-\alpha_2 \rangle$, $|\alpha_1, -\alpha_2 \rangle$, and
$|-\alpha_1, -\alpha_2 \rangle$. This means that quantum
entanglement of the two single photons in the two probe-laser
modes can induce the probabilistic generation of atomic entangled
coherent states for the two distant BECs through projective
measurements.

\begin{figure}[htb]
\begin{center}
\includegraphics[width=8cm,height=2.5cm]{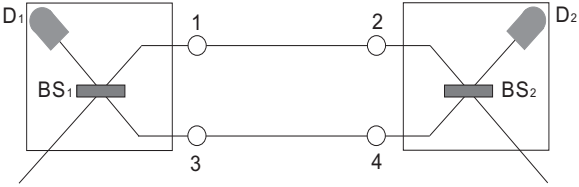}
\end{center}
\vskip 0.2cm \caption{The schematic diagram of creating an
entangled single-photon state given by Eq. (\ref{17}) from
two-pairs of two-mode squeezed vacuum states by using two beam
splitters and two on-off photon detectors. } \label{fig3}
\end{figure}

The preparation of the entangled single-photon state given by Eq.
(\ref{17}) and Bell measurements of optical Bell states $|B
(\pm\pi/4, 0)\rangle$ given by Eq. (\ref{34}) are significant
ingredients in the present single-photon scheme. As for the
preparation of these optical Bell states, Browne \emph{et al.}
\cite{bro,daf} have developed a simple protocol to generate the
entangled single-photon state given by Eq. (\ref{17}) using only
beam splitters and on-off photon detectors from two pairs of
two-mode squeezed vacuum states. The setup of the BEST protocol
\cite{bro} is indicated in Fig. 3 which consists of two beam
splitters, two on-off photon detectors. Modes 1 and 2, modes 3 and
4  are fed with weak two-mode squeezed vacuum states
$(1-q^2)\Sigma_{n,m}q^{n+m}|n,n\rangle_{12}\otimes|m,m\rangle_{34}$,
which can be approximately expressed as the following
un-normalized state (up to the order of $q^2$)
\begin{eqnarray}
\label{43} |\psi_q\rangle_{1234} &\approx&|00 00\rangle +q(|00
11\rangle
+|11 00\rangle)\nonumber\\
&&+q^2(|00 22\rangle)+|11 11\rangle+|22 00\rangle),
\end{eqnarray}
where the parameter $q=\tanh(r)$  ($0\leq q\ll 1$) is determined
by the squeezing parameter $r$. Such states with low values of $r$
are easily produced in experiments. For instance, two-mode
squeezed vacuum states have been used to demonstrate
experimentally quantum teleportation of a coherent state
\cite{fur} and quantum dense coding with continuous variables
\cite{miz}. In these experiments the squeezing parameter takes its
values of $r\simeq 0.35$ and $0.23$, respectively.

Let $\hat{U}_{1}(r_1,t_1)$ and  $\hat{U}_{2}(r_2,t_2)$ denote
unitary transformations of the two beam splitters with
transmissivity $t_1$ and $t_2$,  and reflectivity $r_1$ and $r_2$,
which satisfy the relation $|r_i|^2+|t_i|^2=1$ with $i=1,2$. The
two beam-splitter transformations are given by
\begin{eqnarray}
\label{44} \hat{U}_{1}(r_1,t_1) &=&t_1^{a^{\dag}_1a_1} e^{-r^*_1
a^{\dag}_3a_1} e^{r_1 a_3a^{\dag}_1}
t_1^{-a^{\dag}_3a_3},\nonumber\\
\hat{U}_{2}(r_1,t_2) &=&t_1^{a^{\dag}_2a_2} e^{-r^*_2
a^{\dag}_4a_2} e^{r_2 a_4a^{\dag}_2} t_2^{-a^{\dag}_4a_4},
\end{eqnarray}
which produce the output state of the two beam splitters given by
$\hat{U}_{1}(r_1,t_1)\hat{U}_{2}(r_2,t_2)|\psi_q\rangle_{1234}$
with the following expression
\begin{eqnarray}
\label{45} &&|00\rangle_{13}\times |00\rangle_{24}\nonumber\\
&&+q(r_1|10\rangle +t^{*}_1|01\rangle)_{13}\times(r_2|10\rangle+t^{*}_2|01\rangle)_{24}\nonumber\\
&&+q(t_1|10\rangle-r^{*}_1|01\rangle)_{13}\times(t_2|10\rangle-r^{*}_2|01\rangle)_{24}\nonumber\\
&&+\frac{1}{2}q^2\left(\sqrt{2}r^{2}_1|20\rangle+\sqrt{2}t^{*2}_1|02\rangle+2r_1t^{*}_{1}|11\rangle\right)_{13}\nonumber\\
&&\otimes\left(\sqrt{2}r^{2}_2|20\rangle+\sqrt{2}t^{*2}_2|02\rangle+2r_2t^{*}_{2}|11\rangle\right)_{24}\nonumber\\
&&+q^2\left[\sqrt{2}t_1r_1|2 0\rangle -\sqrt{2}t^{*}_1r{*}_1|0 2\rangle+(1-2|r_1|^2)|11\rangle\right]_{13} \nonumber\\
&&\otimes \left[\sqrt{2}t_2r_2|2 0\rangle -\sqrt{2}t^{*}_2r{*}_2|0
2\rangle +(1-2|r_2|^2)|11\rangle\right]_{24}\nonumber\\
&&+\frac{1}{2}q^2\left[\sqrt{2}t^{2}_1|2 0\rangle -\sqrt{2}r^{*2}_1|0 2\rangle-2t_1r^{*}_1|11\rangle\right]_{13} \nonumber\\
&&\otimes \left[\sqrt{2}t^{2}_2|2
0\rangle-\sqrt{2}r^{*2}_2|02\rangle
-2t_2r^{*}_2|11\rangle\right]_{24}.
\end{eqnarray}
In the derivation of Eq. (\ref{45}) we have used the following
transformation relations for photonic creation operators of the
incoming fields \cite{sche,leon}
\begin{eqnarray}
\label{46} \hat{U}_{1}\left(
\begin{array}{c}
a^{\dag}_1 \\
a^{\dag}_3  \\
\end{array} \right)\hat{U}^{\dag}_{1}&=&\left(
\begin{array}{cc}
t_1&-r^{*}_1\\
r_1&t^{*}_1 \\
\end{array} \right)
\left(
\begin{array}{c}
a^{\dag}_1 \\
a^{\dag}_3  \\
\end{array} \right),\nonumber \\
\hat{U}_{2}\left(
\begin{array}{c}
a^{\dag}_2 \\
a^{\dag}_4  \\
\end{array} \right)\hat{U}^{\dag}_{2}&=&\left(
\begin{array}{cc}
t_2&-r^{*}_2\\
r_2&t^{*}_2 \\
\end{array} \right)
\left(
\begin{array}{c}
a^{\dag}_2 \\
a^{\dag}_4  \\
\end{array} \right).
\end{eqnarray}

The on-off detectors distinguish between vacuum and the presence
of any number photons. An ideal on-off detection with quantum
efficiency $\eta=1$ can be described by the positive
operator-valued measure (POVM) of each detector
$\hat{\Pi}=|0\rangle\langle0|$ and $\hat{D}=I-|0\rangle\langle0|$.
We consider two on-off detectors with the same quantum efficiency
$\eta$ by the operator \cite{oli,sas}
\begin{equation}
\label{47}
\hat{D}_i(\eta)=\sum^{\infty}_{n=0}(1-\eta)^{n}|n\rangle_i|\langle
n|, \hspace{1cm} (i=1,2).
\end{equation}

Then the entangled single-photon state of the form defined by Eq.
(\ref{17}) can be produced  for arbitrary values of the parameters
$\theta$ and $\varphi$ in Eq. (\ref{17}) through the procedure
indicated in Fig. 3 with the following output state
\begin{equation}
\label{48}
|\Psi\rangle_{1234}=\hat{D}_1(\eta)\hat{D}_2(\eta)\hat{U}_{1}(r_1,t_1)\hat{U}_{2}(r_2,t_2)|\psi_q\rangle_{1234},
\end{equation}
which describes two identical two-mode squeezed states are mixed
pairwise at unbalanced beam splitters followed by the on-off
photon detections which are indicated in Fig. 3.

Through length but straightforward calculations, making use of
Eqs. (\ref{45}) and (\ref{47}) from Eqs. (\ref{43}) and (\ref{48})
we can find that
\begin{eqnarray}
\label{49} |\Psi\rangle_{1234}&=&q\chi_1|00 11\rangle +
q^2[\chi_2|11 11\rangle+\chi_3|10 12\rangle \nonumber\\
&&+\chi_4|00 22\rangle+\chi_5|01 21\rangle],
\end{eqnarray}
where the five coefficients are given by
\begin{eqnarray}
\label{50} \chi_1&=&(r^*_1r^*+t^*_1t^*_2)\eta^2,
\nonumber\\
\chi_2&=&[2r_1r_2t^*_1t^*_2
+2t_1t_2r^*_1r^*_2+(1-2|r_1|^2)(1-2|r_2|^2)]\eta^2,
\nonumber\\
\chi_3&=&\sqrt{2}[r_1t^*_1t^{*2}_2-r^*_1t^*_1r^{*2}_2-r^*_2t^*_2(1-2|r_1|^2)]\eta^2(2-\eta), \nonumber\\
\chi_5&=&\sqrt{2}\left[r_2t^{*}_2t^{*2}_1
-t_2r^{*}_2r^{*2}_1-r^{*}_1t^{*}_1(1-2|r_2|^2)\right]\eta^2(2-\eta),\nonumber\\
\chi_4&=&(r^{*}_1r^{*}_2-t^*_1t^*_2)^2\eta^2(2-\eta).
\end{eqnarray}

After taking trace of the state (\ref{49}) over modes 3 and 4, we
find the reduced density operator of the normalized state of modes
1 and 2 to be the following form
\begin{eqnarray}
\label{51} \rho_{12}&=&P_{\psi}|\psi\rangle_{12}\langle\psi|
+P_{10}(|10\rangle_{12}\langle10| \nonumber\\
&&+P_{00}(|00\rangle_{12}\langle00|
+P_{01}(|01\rangle_{12}\langle01|,
\end{eqnarray}
where the producing probabilities of the corresponding states are
given by
\begin{eqnarray}
\label{52}P_{\psi}&=&\frac{|\chi_1|^2+q^2|\chi_2|^2}{\Sigma},
\hspace{0.3cm} P_{10}=\frac{q^2|\chi_3|^2}{\Sigma}, \nonumber\\
P_{00}&=&\frac{q^2|\chi_4|^2}{\Sigma}, \hspace{0.3cm}
P_{01}=\frac{q^2|\chi_5|^2}{\Sigma},
\end{eqnarray}
where we have introduced the symbol
\begin{equation}
\label{53}\Sigma=|\chi_1|^2+q^2(|\chi_2|^2+|\chi_3|^2+|\chi_4|^2+|\chi_5|^2),
\end{equation}
and the expected state $|\psi\rangle_{12}$ has the following form
\begin{eqnarray}
\label{54}
|\psi\rangle_{12}=\frac{\chi_1|00\rangle_{12}+q\chi_2|11\rangle_{12}}{\sqrt{|\chi_1|^2+q^2|\chi_2|^2}},
\end{eqnarray}
which is the entangled single-photon state of the form defined by
Eq. (\ref{17}). From the expressions of $\chi_1$ and $\chi_2$
given by Eq. (\ref{50}) we can see that the entangled
single-photon state by Eq. (\ref{54}) is independent of the
quantum efficiency of the on-off detectors. This means that the
quantum efficiency of the on-off detectors does not affect the
fidelity of the prepared state.

From Eqs. (\ref{51}-\ref{54}) we can see that the entangled
single-photon state (\ref{17}) or (\ref{54})  can be
near-deterministically produced if $P_{\psi}\approx 1$,
$P_{10}\approx 0$, $P_{00}\approx 0$, and $P_{01}\approx 0$. In
what follows we show that two optical Bell states defined by Eq.
(\ref{34}) can be obtained using above procedure through properly
choosing parameters of the two beam splitters. In order to see
this, we choose $t_2\approx 0$ and $r_2\approx 1$ for the second
beam splitter, and take $q\chi_2=\pm\chi_1$ which leads to
$r_1\approx \pm q$ and $t_1\approx \sqrt{1-q^2}$ for the first
beam splitter in which $q$ is defined in  Eq. (\ref{43}).  In this
case, the five coefficients given by Eq. (\ref{50}) become
\begin{eqnarray}
\label{55} \chi_1&=&\pm q\eta^2, \chi_2=\pm q^{-1}\chi_1,
\chi_4=q^2\eta^2(2-\eta),  \nonumber\\
\chi_3&=&-\chi_5=\mp q\sqrt{2(1-q^2}\eta^2(2-\eta),
\end{eqnarray}
and the entangled single-photon state $|\psi\rangle_{12}$ becomes
\begin{eqnarray}
\label{56} |\psi\rangle_{12}=\frac{1}{\sqrt{2}}(|00\rangle_{12}\pm
|11\rangle_{12}).
\end{eqnarray}
which are two optical Bell states defined in  Eq. (\ref{34}).  And
the success probabilities of the corresponding states in Eq.
(\ref{52}) become
\begin{eqnarray}
\label{57}P_{\psi}&=&\frac{2}{2+q^2(4-3q^2)(2-\eta)^2},\nonumber\\
P_{01}&=&P_{10}=\frac{2q^2(1-q^2)(2-\eta)^2}{2+q^2(4-3q^2)(2-\eta)^2},\nonumber\\
P_{00}&=&\frac{q^4(2-\eta)^2}{2+q^2(4-3q^2)(2-\eta)^2},
\end{eqnarray}
which indicate that in the weak squeezing regime of $0<q\ll 1$  we
can always find  $P_{\psi}\approx 1$, $P_{10}\approx 0$,
$P_{\psi}\approx 0$, and $P_{01}\approx 0$ due to  $0<\eta \leq
1$. In fact, from the expressions of the probabilities given by
Eqs. (\ref{57}) we can see that the quantum efficiency of the two
detectors does not seriously  affect  the success probability of
generating optical Bell states defined in Eq. (\ref{17}). For
instance, when $q=0.1$ and $\eta=1$, we have $P_{\psi}\approx
98.98\%$  while when $q=0.1$ and $\eta=0.7$, we have
$P_{\psi}\approx 98.97\%$. Therefore, in the scheme under our
consideration optical Bell states (\ref{34}) can be produced
perfectly with the success probability of approaching unity by
using two pairs of two-mode squeezed vacuum states in the weak
squeezing regime and on-off detections of photons.

It is a nontrivial thing to distinguish two optical Bell states
defined by Eq. (\ref{34}), which are energy-entangled Bell states
since they are defined in the energy representation. To our
knowledge, there are a few methods for Bell measurements of the
polarization-entangled Bell states and the momentum-entangled Bell
states of photons \cite{sch,wal}, but no scheme is available for
Bell measurements of the energy-entangled Bell states of photons.
Recent achievements on cavity QED \cite{rai} provide us with the
possibility of distinguishing energy-entangled Bell states of
photons. In what follows we show that two optical Bell states
defined by Eq. (\ref{34}) can be distinguished using entanglement
in additional auxiliary atomic systems through mapping two optical
Bell states into two atomic Bell states. In order to do this,
consider two single-mode cavities which are initially fed with one
of the two energy-entangled Bell states given by Eq. (\ref{34}).
Each cavity contains a two-level atom. Two atoms in two cavities
interact resonantly with two cavity fields with the Hamiltonian
\begin{eqnarray}
\label{58}
H'=\hbar\sum^{2}_{r=1}\left[\frac{\omega_i}{2}\sigma^{r}_z
+\omega_r a^{\dag}_r a_r - i\frac{\Omega}{2}(a_r\sigma^{r}_+
-\sigma^{r}_-a^{\dag}_r)\right],
\end{eqnarray}
which implies that if the two cavity fields are initially in two
optical Bell states given by Eq. (\ref{34}) and two atoms are
initially in the ground state $|g, g\rangle$, then at the time of
$\Omega t=\pi$, the state of the field-atom system governed by Eq.
(\ref{58}) becomes
\begin{eqnarray}
\label{59} |\Psi(\pi)\rangle&=&\frac{1}{\sqrt{2}}U(\Omega
t=\pi)(|0,0\rangle \pm |1,1\rangle)|g, g\rangle\nonumber\\
&=&\frac{1}{\sqrt{2}}|0,0\rangle(|g,g\rangle \pm |e, e\rangle),
\end{eqnarray}
where $U$ is the time evolution operator corresponding to the
Hamiltonian (\ref{58}).

From Eq. (\ref{59}) we can see that the discrimination of the two
optical Bell states given by Eq. (\ref{34}) is now changed into
distinguishing two atomic Bell states $(|g,g\rangle \pm |e,
e\rangle)/\sqrt{2}$. Assume that the two atoms in the two cavities
are initially in the two atomic Bell stats $(|g,g\rangle \pm |e,
e\rangle)/\sqrt{2}$, applying two classical laser fields to the
two atoms, it is easy to find that the two atomic Bell states with
the $\pi/2$ evolution of the system performs the following
transformations
\begin{eqnarray}
\label{60} \frac{1}{\sqrt{2}}(|g,g\rangle + |e,
e\rangle)&\rightarrow&\frac{1}{\sqrt{2}}(|e,e\rangle + |g,
g\rangle)\nonumber\\
\frac{1}{\sqrt{2}}(|g,g\rangle - |e,
e\rangle)&\rightarrow&\frac{1}{\sqrt{2}}(|e,g\rangle - |g,
e\rangle),
\end{eqnarray}
which indicate that when both of the two atoms in the two cavities
are detected to be in the excited state $|e, e\rangle$ or the
ground state $|g, g\rangle$, the atomic Bell state  $(|g,g\rangle
+ |e, e\rangle)/\sqrt{2}$ is identified. When one atom is detected
to be in the excited state and another to be in the ground state,
i.e., $ |e, g\rangle$ or $ |g, e\rangle$, the atomic Bell state
$(|g,g\rangle - |e, e\rangle)/\sqrt{2}$ is identified. This
implies the discrimination of the two energy-entangled Bell states
of photons given by Eq. (\ref{34}).

It is worthwhile to mention that the physical mechanism of
generating atom-atom entangled states between two distant BECs in
this section is very different from that of producing atom-photon
entangled states in the previous section. There are two key
factors to generate quantum entanglement between two distant BECs.
One is that the two probe laser fields must have the initial
quantum entanglement as indicated in the above discussion. In this
sense the appearance of the entanglement of the two distant BECs
is the result of entanglement transferring from the two probe
lasers to the two distant BECs  through performing optical Bell
measurements. The other is that the concrete form of entangled
coherent states depends upon the parameters of the applied lasers
and the two BECs and the interaction time.

It should be mentioned that one drawback of using the entangled
single-photon state as a quantum channel for generating entangled
coherent states of distant BECs is that it is extremely sensitive
to photon loss and detection inefficiency. In particular, the
completely correlated state is destroyed by loss of even a single
photon. Other correlated states may have somewhat less strict
requirements. In order to overcome this disadvantage in the
following we will use the entangled coherent-state channel to
create entangled coherent states of distant BECs. Because
entangled coherent states are multiphoton states, they are more
robust against small levels of photon absorption losses.

\begin{center}
{\bf B. The entangled coherent-state scheme}
\end{center}

In complete analogy to the above discussion, the entangled
coherent state channel can be used to create entangled states of
distant BECs. In order to see this, we assume that the two distant
BECs are initially uncorrelated and they are in a two-mode
coherent state $|\alpha_1, \alpha_2\rangle$ while the pair of the
probe lasers are initially in the entangled coherent state $|E
(\beta,\beta)\rangle_+$ defined by Eq. (\ref{35}) which is a
quasi-Bell basis. As well known, for a given nonzero number
$\beta$ there are four linearly independent two-mode coherent
states $|\beta, \beta\rangle$, $|\beta, -\beta\rangle$, $|-\beta,
\beta\rangle$, and $|-\beta, -\beta\rangle$. They compose four
normalized quasi-Bell basis defined by
\begin{eqnarray}
\label{61} |qB (\beta, \beta)\rangle_{\pm}&=&N_{\pm}(\beta)||qB
(\beta, \beta)\rangle,
\nonumber \\
|qB (\beta, -\beta)\rangle_{\pm}&=&N_{\pm}(\beta)||qB (\beta,
-\beta)\rangle,
\end{eqnarray}
where the normalization constants
$N^{-1/2}_{\pm}(\beta)=2\pm2\exp(-4|\beta|^2)$ and four
un-normalized quasi-Bell states are given by
\begin{eqnarray}
\label{62} ||qB (\beta, \beta)\rangle_{\pm}&=&|\beta, \beta\rangle
\pm |-\beta, -\beta\rangle,
\nonumber \\
||qB (\beta, -\beta)\rangle_{\pm}&=&|\beta, -\beta)\rangle \pm
|-\beta,\beta\rangle.
\end{eqnarray}

The four normalized quasi-Bell states defined by Eq. (\ref{61})
are orthogonal to each other except $|qB (\beta,\beta)\rangle_+$
and $|qB (\beta,-\beta)\rangle_+$ which have the following
orthogonal relation
\begin{eqnarray} \label{63}
_+\langle qB (\beta,\beta)|qB
(\beta,-\beta)\rangle_+=\frac{1}{\cosh(2|\beta|^2)},
\end{eqnarray}
which indicates that the two quasi-Bell states $|qB
(\beta,\beta)\rangle_+$ and $|qB (\beta,-\beta)\rangle_+$ rapidly
become orthogonal as $\beta$ grows. For example, when $|\beta|\geq
2$, the overlap $ _+\langle qB (\beta,\beta)|qB
(\beta,-\beta)\rangle_+\leq 2.2\times 10^{-7}$.

Obviously, these quasi-Bell basis are a type of special entangled
coherent states. Since an entangled coherent state is a
superposition state of all number states of photons for the two
probe laser modes, the scheme based on the entangled
coherent-state quantum channel is actually a multiphoton protocol.
Then the joint initial state of the whole atom-photon system is
given by
\begin{equation}
\label{64} |\Phi(0)\rangle =\frac{1}{\sqrt{N_{\beta
\beta}}}(|\beta,\beta\rangle +
|-\beta,-\beta\rangle)\otimes|\alpha_1, \alpha_2\rangle.
\end{equation}

Assume that the lasers are applied adiabatically at the time of
$t=0$, then from Eqs. (\ref{13}), (\ref{16}) and (\ref{18}) it is
straightforward to show that the joint state of the system after
the interaction time $t$ becomes a superposition state of two
wavefunction given by
\begin{equation}
\label{65} |\Phi(t)\rangle =\frac{1}{\sqrt{N_{\beta
\beta}}}(|\varphi_{\beta \alpha}(t)\rangle + |\varphi_{-\beta
\alpha}(t)\rangle),
\end{equation}
where the two wave functions $|\varphi_{\pm\beta \alpha}(t)\rangle$
are defined by
\begin{eqnarray}
\label{66} |\varphi_{\pm\beta \alpha}(t)\rangle
&=&\prod^2_{n=1}|\varphi_{\pm\beta \alpha}(t)\rangle_n,
\end{eqnarray}
where we have introduced
\begin{equation}
\label{67} |\varphi_{\pm\beta \alpha}(t)\rangle_n=
 \exp(-it\hat{H}_n)|\pm\beta\rangle_n|\alpha_n\rangle,
\end{equation}
where $\hat{H}_n$ denotes the effective interaction Hamiltonian
between $n$th probe laser and $n$th BEC ($n=1,2$) given by
\begin{equation}
\label{68} \hat{H}_n=\omega'_{n}\hat{b}^{\dagger}_{n}\hat{b}_{n}
+2\omega'_{n}\hat{b}^{\dagger}_{n}\hat{b}_{n}\hat{a}^{\dagger}_n\hat{a}_n+\lambda_n
\hat{b}^{\dagger 2}_{n}\hat{b}^2_{n}.
\end{equation}

In a previous paper \cite{kua}, it has bee shown that the states
given by Eq.  (\ref{67}) are two-mode generalized coherent states.
They can be expressed as continuous superposition states of
two-mode product coherent states
\begin{equation}
\label{69}  |\varphi_{\pm\beta \alpha}(\tau)\rangle_n =
\frac{1}{(2\pi)^2}\int^{2\pi}_0d\phi_1d\phi_2
 f(\phi_1, \phi_2)\left|\pm\beta e^{i\phi_1}, \alpha_n e^{i\phi_2}\right\rangle,
\end{equation}
where we have set $\omega'_{1}=\omega'_{2}$,
$\lambda'_{1}=\lambda'_{2}=\lambda$, and  introduce a scaled time
$\tau=\lambda t$. The phase function in Eq. (\ref{69}) is given by
\begin{eqnarray}
\label{70}
 f(\phi_1, \phi_2)&=&\exp\left[i\left(\tau\theta_{n,m} - n\phi_1
- m\phi_2\right)\right], \\
\label{71}
 \theta_{n,m}&=&(1+K)m+2Knm-m^2,
\end{eqnarray}
where the effective interaction parameter $K$ is defined in Eq.
(\ref{24}).

From Eqs.  (\ref{69}-\ref{71}) we can see that when $K$ takes an
integer, $|\varphi_{\pm\beta \alpha}(\tau)\rangle_n$ is periodic
with the period of $2\pi$. In particular, when the scaled time
$\tau$ takes its values in the following  manner
\begin{equation}
\label{72} \tau= \frac{M}{N}2\pi,
\end{equation}
where $M$ and $N$ are mutually prime integers, we can find the
states given by  Eq. (\ref{69}) become discrete superposition
states of product coherent states
\begin{equation}
\label{73} |\varphi_{\pm\beta \alpha}(\tau=2\pi M/N)\rangle_n =
\sum^{N}_{r=1} \sum^{N}_{s=1}c_{rs}\left|\pm\beta e^{i\varphi_r},
\alpha_n e^{i\varphi_s} \right\rangle,
\end{equation}
where the two running phases are defined by
\begin{equation}
\label{74} \varphi_r=\frac{2\pi}{N}r, \hspace{0.5cm}
\varphi_s=\frac{2\pi}{N}s, \hspace{0.5cm} (r, s=1,2, \cdots, N),
\end{equation}
and the superposition coefficients $c_{r s}$ to be given by
\begin{equation}
\label{75} c_{rs}=\frac{1}{N^2}\sum^{N}_{n,m=1}
\exp\left\{-\frac{2\pi i }{N}\left[nr + ms -M \theta_{n,
m}\right]\right\}.
\end{equation}

Then after an interaction time $\tau=2\pi M/N$ with the lasers the
joint state given by Eq. (\ref{65}) $|\Phi(\tau=2\pi M/N)\rangle$
becomes
\begin{widetext}
\begin{eqnarray}
\label{76}\left|\Phi\left(\tau=\frac{M}{N}2\pi
\right)\right\rangle&=&\frac{1}{\sqrt{N_{\beta
\beta}}}\sum^{N}_{r, s=1}
 \sum^{N}_{r', s'=1}c_{r s}c_{r' s'}[|\beta e^{i\varphi_r}, \beta
 e^{i\varphi_{r'}}\rangle+ |-\beta e^{i\varphi_r}, -\beta
 e^{i\varphi_{r'}}\rangle]\otimes|\alpha_1 e^{i\varphi_s}, \alpha_2
 e^{i\varphi_{s'}}\rangle,
\end{eqnarray}
\end{widetext}
which is generally an atom-photon entangled coherent state between
the two probe laser fields and the two distant BECs. We note that
the state given by Eq. (\ref{76}) can be expressed as a
superposition state of un-normalized two-mode Cat states of the
probe laser fields and product coherent states of the two distant
BECs in the following simple form
\begin{eqnarray}
\label{77} &&\sum^{N}_{r, s=1}
 \sum^{N}_{r', s'=1}c_{r s}c_{r' s'}||C(\beta e^{i\varphi_r}, \beta
 e^{i\varphi_{r'}})\rangle  \nonumber \\
 & &\otimes|\alpha_1 e^{i\varphi_s}, \alpha_2
 e^{i\varphi_{s'}}\rangle,
\end{eqnarray}
where the un-normalized two-mode Cat states are defined by
\begin{equation}
\label{78}||C(\beta e^{i\varphi_r}, \beta
 e^{i\varphi_{r'}})\rangle=|\beta e^{i\varphi_r},
\beta e^{i\varphi_{r'}}\rangle+|-\beta e^{i\varphi_r}, -\beta
 e^{i\varphi_{r'}}\rangle.
\end{equation}

In general, two Cat states defined by Eq. (\ref{78})are not
orthogonal with each other. The Cat states appeared in the
atom-photon entangled state given by Eq. (\ref{77})  have the same
amplitudes but different phases, their orthogonal relation is
given by
\begin{widetext}
\begin{equation}
\label{79}\langle C(\beta e^{i\varphi_r}, \beta e^{i\varphi_{r'}})
||C(\beta e^{i\varphi_s}, \beta
e^{i\varphi_{s'}})\rangle=4e^{-2|\beta|^2}\cosh\left\{|\beta|^2\left[\exp\left(\frac{2\pi
i}{N}(s-r)\right)+ \exp\left(\frac{2\pi
i}{N}(s'-r')\right)\right]\right\},
\end{equation}
\end{widetext}
which implies that one can not obtain entangled coherent states of
the two distant BECs from the atom-photon entangled state given by
Eq. (\ref{76}) through performing projective measurements on these
Cat states.

Without loss of generality, but avoiding the lengthy calculation
relative to the state given by Eq. (\ref{76}), as a specific
example, let us have a look at the case of $\tau=\pi/2$ with
$K=-1$, $N=4$, and $M=1$. In this case, we find the joint state of
the system given by Eq. (\ref{76}) becomes
\begin{eqnarray}
\label{80} \left|\Phi(\tau=\pi/2)\right\rangle
&=&\frac{1}{4}\left[ |qB (\beta,\beta)\rangle_+\otimes||E(\alpha_1,-\alpha_2)\rangle \right. \nonumber \\
& &\left.- i|qB
(\beta,-\beta)\rangle_+\otimes||E(\alpha_1,\alpha_2)\rangle_{-}
\right],
\end{eqnarray}
where $||E(\alpha_1,\alpha_2)\rangle_{-}$ is an entangled coherent
state defined by
\begin{equation}
\label{81}  ||E(\alpha_1, \alpha_2)\rangle_-= |\alpha_1,
\alpha_2\rangle - |-\alpha_1,-\alpha_2\rangle,
\end{equation}
which reduces to a quasi-Bell state when $\alpha_2=\pm\alpha_1$.

From the atom-photon entangled state given by Eq. (\ref{80}) we
can see that entangled coherent states of the two distant BECs can
be produced through performing nonlocal Bell measurements on
quasi-Bell states of the two probe lasers. When the two probe
lasers are in quasi-Bell states $|qB (\beta,\beta)\rangle_+$ and
$|qB (\beta,-\beta)\rangle_+$, the two distant BECs collapses to
entangled coherent states denoted by un-normalized entangled
coherent states $||E(\alpha_1,-\alpha_2)\rangle$ and
$||E(\alpha_1,\alpha_2)\rangle_{-} $, respectively. After
normalization the first  entangled coherent state denoted by
$|E(\alpha_1,-\alpha_2)\rangle$ is the same as that given by Eq.
(\ref{42}) in the single-photon protocol. It is interesting to
note that the state given by Eq. (\ref{73}) is an atom-photon
entangled state between a pair of optical quasi-Bell states of the
two probe lasers and a pair of atomic entangled coherent states of
the two distant BECs. This kind of atom-photon entangled states
gives a new entanglement resource for quantum networks based on
the atom-photon quantum interface.

It should be pointed out that in the entangled coherent-state scheme we have to prepare the
quasi-Bell state $|qB (\beta,\beta)\rangle_+$ and to perform Bell measurements with respect to four
quasi-Bell basis represented by $|qB (\beta,\beta)\rangle_{\pm}$ and $|qB
(\beta,-\beta)\rangle_{\pm}$. It is a nontrivial matter to prepare a quasi-Bell state. Glancy
\emph{et al.} \cite{gla} proposed a scheme to generate optical quasi-Bell states using a
single-mode cat state. They indicate that the optical quasi-Bell state $|qB (\beta,\beta)\rangle$
can be created by sending a cat state of the form $|-\sqrt{2}\beta\rangle+|\sqrt{2}\beta\rangle$
and the vacuum state into the input ports of a 50:50 beam splitter. Then the output ports of the
beam splitter will contain entangled modes of the optical quasi-Bell state. So the problem of
generating optical quasi-Bell states is reduced to the generation of single-mode optical cat states
while cat states can be created by sending a coherent state into a nonlinear medium exhibiting the
Kerr effect \cite{yur}. Howell and Yeazell \cite{how} presented an alternative approach to the
generation of the optical quasi-Bell states. The setup considered in Ref. \cite{how} applies a
Mach-Zehnder interferometer equipped with a cross-Kerr medium in each of two spatially separated
modes.

Nonlocal Bell measurements with respect to four optical quasi-Bell
basis can be accomplished by the scheme developed in Refs.
\cite{enk,jeo1} with arbitrarily high precision using beam
splitters, phase shifts, nonlinear Kerr medium, coherent light
sources, and photodetectors. In order to perform a Bell
measurement of a optical quasi-Bell state, send the two modes
labelled by $\hat{a}_1, \hat{a}_2$ involved in the corresponding
quasi-Bell state into a 50:50 beam splitter $B_{1
2}=\exp[\pi/4(\hat{a}_1\hat{a}^{\dag}_2-\hat{a}_2\hat{a}^{\dag}_1)]$.
Then use photon counters to measure the number of photons in each
mode. A complete Bell measurement of the quasi-Bell state can be
accomplished through measuring the photon number parity,  the
absence and presence of photons (i.e., zero or nonzero photons) in
the two modes using on-off photon detectors. If an odd (even)
number of photons in mode $\hat{a}_1$ are detected while zero
photons in mode $\hat{a}_2$ is detected, a Bell measurement with
respect to the quasi-Bell state $|qB (\beta,\beta)\rangle_+$ ($|qB
(\beta,\beta)\rangle_-$) is completed. When no photon is detected
in mode $\hat{a}_1$ while even (odd) number of photons in mode
$\hat{a}_2$ are detected, a Bell measurement with respect to the
quasi-Bell state $|qB (\beta,-\beta)\rangle_+$ ($|qB
(\beta,-\beta)\rangle_-$) is done.

From the above discussions we can see that Kerr medium is required
for both preparing and measuring the optical quasi-Bell states in
all of the above schemes.  It is a greater challenge to
experimentally produce large Kerr nonlinearities. Although
sufficiently large Kerr nonlinearities have been difficult to
produce, significant progress is being made in this area. In
particular, recent progress on atomic quantum coherence
\cite{pat,hau,lm,kan,wan} indicates that it is possible to prepare
Kerr media with the giant Kerr nonlinearities through using the
EIT technology. Paternostro and coworkers \cite{pat} proposed a
double EIT scheme which can enhance the cross-Kerr effect in a
dense atomic medium in the EIT regime. Especially, it has been
proved that the interaction of two travelling fields of light in
an atomic medium is able to show giant Kerr nonlinearities by
means of the so-called cross-phase-modulation. Measured values of
the $\chi^{3}$ parameter are up to six orders of magnitude larger
than usual \cite{hau}. Therefore, our protocol is at the reach of
current experiment.

\section{Concluding Remarks}

In this paper, we have presented a method to generate entangled coherent states between two distant
BECs using EIT techniques. Concretely we have proposed two protocols of generating entangled
coherent states of two distant BECs in terms of involved different quantum channels. One is the
single-photon protocol, another is the multiphoton one. Two strong coupling laser beams and two
entangled probe laser beams are used to make two distant BECs be in EIT states and to generate an
atom-photon entangled state between a pair of probe lasers and two distant BECs. The atom-photon
entangled state is a supply of generating entangled coherent states of distant BECs. The concrete
form of the atom-photon entangled state depends on the parameters of applied lasers and BECs, the
initial states of the entangled probe lasers and two BECs, and the interaction time. Since what we
want is to generate entangled coherent states of two distant BECs, we choose the initial states of
two BECs are coherent states. The initial entanglement of the two probe lasers plays a key role in
the proposed method of generating entangled coherent states of two distant BECs. If there is no the
initial entanglement of the two probe lasers, one can not create quantum entanglement of the two
distant BECs in the system under our consideration. In this sense we can say entanglement between
two probe lasers is transferred on two distant BECs via projective measurements. In other words,
the entangled probe lasers induce the generation of entangled distant BECs. The concrete form of
the initial entangled states of the two probe lasers affects the forms of the generated atom-photon
entangled states and projective measurements.  An initial entangled state of the two probe lasers
can be considered as a quantum channel to create entangled states of two distant BECs. Quantum
entanglement is transferred from the two probe lasers to the two distant BECs through performing
projective measurements. We have explicitly shown how to generate entangled coherent states for the
two distant BECs in both of the single-photon and multi-photon protocols.

We have also obtained some atom-photon entangled states of
particular interest between two probe lasers and two distant BECs.
For instance, in the single-photon scheme we have
deterministically generated the atom-photon entangled state
consisting of a pair of optical Bell states and a pair of atomic
quasi-Bell states for the two correlated probe lasers and the two
distant BECs. This entangled state given by Eq. (\ref{33}) is a
hybrid entangled state between discrete-variable states and
continuous variable states, such a kind of hybrid entangled states
can serve as a valuable resource in quantum information processing
and build up a bridge between quantum information protocols of
discrete and continuous variables. In the multi-photon scheme we
have deterministically generated the atom-photon entangled state
consisting of a pair of optical quasi-Bell states and a pair of
atomic quasi-Bell states. These entangled states consisting of
optical Bell or quasi-Bell states and atomic quasi-Bell states
provide us with a new resources of entanglement, and may open new
ways to carry out quantum information processing based on the
atom-photon quantum interface.

We have shown how it is possible to prepare entangled coherent
states of the two distant BECs in both suggested protocols. In
particular, in the single-photon scheme only low two-mode squeezed
vacuum states and quadrature-phase homodyne measurements are
required in the preparation and detection of the optical Bell
states. These make the single-photon protocol experimentally
feasible at present.  In the multiphoton scheme, the hard tasks
are to prepare and detect an optical entangled coherent state. We
have seen that both preparing and measuring an optical entangled
coherent state require nonlinear Kerr media. EIT has been studied
as a method to obtain giant Kerr nonlinearity. There have been
experimental reports of indirect and direct measurements of  giant
Kerr nonlinearities  utilizing EIT \cite{hau,kan,wan}. Although
this developing technology has not been exactly at hand to
generate and to detect an optical entangled coherent state, it
enables the multiphoton protocol at the reach of the present-day
techniques. The experimental realization for the protocols
proposed in the present paper deserves further investigation.

\acknowledgments

This work was supported by the Alexander von Humboldt Foundation,
the National Natural Science Foundation of China (Grant Nos.
10325523 and 10775048), the National Fundamental Research Program
of China (Grant No. 2007CB925204), and the Education Department of
Hunan Province.

\end{document}